\documentclass[runningheads]{llncs}

\usepackage[T1]{fontenc}
\usepackage{graphicx}

\usepackage{xspace}
\usepackage{flushend}
\usepackage[normalem]{ulem}
\usepackage{array}
\usepackage{multirow}
\usepackage{tikz}
\usepackage{booktabs}

\long\def\ignore#1{}
\sloppypar

\newcommand{\ApproxSign}{\raise.17ex\hbox{$\scriptstyle\sim$}}

\newcommand{\putsec}[2]{\vspace{-0.0in}\section{#2}\label{sec:#1}\vspace{-0.0in}}
\newcommand{\putssec}[2]{\vspace{-0.0in}\subsection{#2}\label{ssec:#1}\vspace{-0.0in}}

\newcommand{\tabref}[1]{Table~\ref{#1}}
\newcommand{\figref}[1]{Figure~\ref{#1}}
\newcommand{\secref}[1]{Section~\ref{sec:#1}}
\newcommand{\ssecref}[1]{Section~\ref{ssec:#1}}

\newcommand{\PNAME}{\mbox{CompPow}\xspace}
\newcommand{\PNAMEBOLD}{\mbox{\textbf{CompPow}}\xspace}

\begin{document}

\title{\PNAME: A Case for Component-level \\GPU Power Management}
\author{Shaizeen Aga
\and
Mohamed Assem Ibrahim
}
\authorrunning{S. Aga, M. Ibrahim}

\institute{Advanced Micro Devices, Inc.\\
\email{\{shaizeen.aga, mohamed1.ibrahim\}@amd.com}}

\maketitle              

\begin{abstract}
The ever increasing demand for ML-driven intelligence in a wide spectrum of domains has led to ubiquity of GPUs. At the same time, GPUs are notorious for their power consumption needs and often dominate power allocation in a typical ML datacenter. While datacenter-level power optimizations which focus on collection of GPUs are promising, in this work, we take a different tack - namely, we take a closer look at power consumption inside a GPU. Specifically, as modern GPUs are comprised of integrated components, we make a case for component-awareness, termed \PNAMEBOLD in this work, for improved power management in modern GPUs. We demonstrate for a variety of ML operations and execution patterns, \PNAME has the potential to deliver higher energy efficiency (10\%) and even improved performance (5\%). We conclude with recommendations on how component-aware software-hardware co-design can extract additional energy efficiency from modern GPUs.

\end{abstract}

%%%%%% -- PAPER CONTENT STARTS-- %%%%%%%%
\putsec{intro}{Introduction}

\begin{figure}[t]
    \centering
    \includegraphics[width=\linewidth]{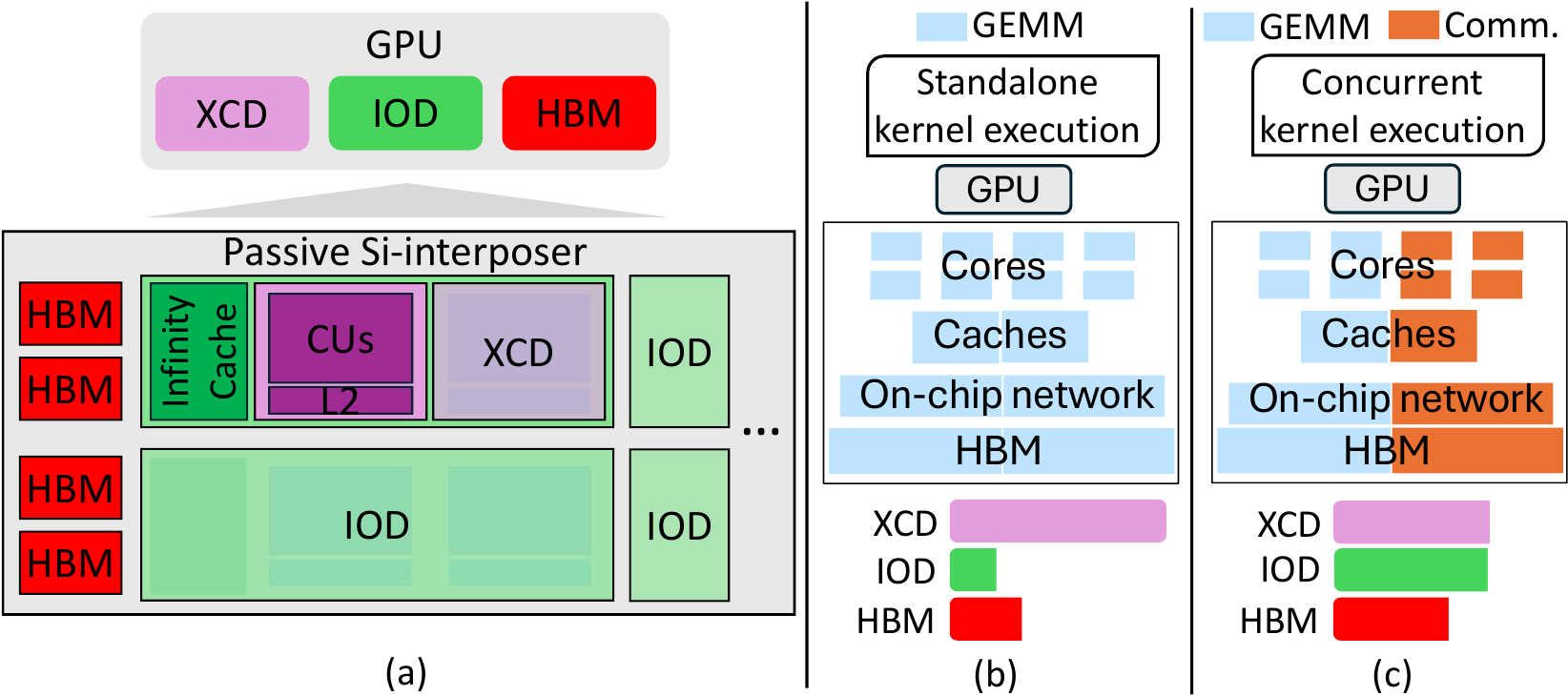}
    \caption{(a) Component-level view of MI300X GPU. (b) Example component-level signature for standalone kernel execution. (c) Example component-level signature for concurrent kernel execution.}
    \label{fig:intro_mi300x_comp_signatures}
    \vspace{-\baselineskip}
\end{figure}

Increasing demand for ML-driven intelligence across a spectrum of domains (healthcare, finance, software development, etc.) is fueled by clusters of GPUs. Electricity demand for such ML-focused datacenters is projected to more than quadruple by 2030~\cite{ieaReport}, with more than half of power allocation therein being to GPUs. As such, there have been prior works~\cite{litSilicon2025,perseusChung24} which have rightly focused on cluster-level power optimizations. While these solutions are valuable, in this work we take a separate tack: we focus on power allocation inside a single GPU and optimizations possible therein. 

Specifically, as we depict in \figref{fig:intro_mi300x_comp_signatures}a, a state-of-the-art GPU such as AMD Instinct\textsuperscript{\texttrademark} MI300X, is comprised of multiple components. That is, a chiplet-based design integrates accelerator complex dies (XCD) which are vertically stacked over I/O dies (IOD), which in turn are stacked over a passive silicon interposer.\footnote{Other GPUs~\cite{blackwellMultiDie} also manifest such integration.} While the XCDs house the GPU compute cores (compute units or CUs) and L2 cache, the IOD houses the LLC (AMD Infinity Cache\textsuperscript{\texttrademark}) and memory interface to the on-package eight stacks of high-bandwidth memory (HBM). 

We show in this work how different operations stress these three key components (XCD, IOD, HBM) differently. As an example, a standalone execution of a general matrix multiplication/GEMM kernel could stress XCD fully as shown in \figref{fig:intro_mi300x_comp_signatures}b, but when run concurrently with a communication collective kernel, both XCD and IOD components could be stressed (\figref{fig:intro_mi300x_comp_signatures}c). Note that, \figref{fig:intro_mi300x_comp_signatures} only pictorially depicts possible (not actual measured) power profiles. That said, we evaluate and discuss measured standalone and concurrency power profiles in this work. We further show how studying component-level power consumption opens up an interesting space for both power sloshing and power optimization within a GPU.

To this end, we make a case in this work for component-awareness for improved power management in modern GPUs. As modern GPUs are often composed of distinct components, a more fine-grain component-level power management design from software to hardware stands a better chance to both extract additional energy efficiency and potentially better performance control. We study key ML operators (general matrix multiplication/GEMM, communication collectives), under standalone and concurrent execution scenarios and show how component-awareness, which we term \PNAMEBOLD, stands to deliver up to 10\% higher energy efficiency and up to 5\% performance. Overall, we believe our work makes a strong case for component-awareness in power management and we conclude with software and hardware recommendations in this regard. 

This work makes the following key contributions:
\begin{enumerate}
\item We make a case in this work for component-awareness at both software and hardware levels for improved power management in modern GPUs. 
\item We demonstrate how component-awareness can deliver higher energy efficiency for communication collectives in ML, a key bottleneck in large-scale ML training and inference.
\item We also demonstrate that for matrix-matrix multiplications (GEMMs), a key work-horse for the pervasive ML workloads, different components in the GPU jostle for power allocation. We further show how component-level utilization metrics can help guide fine-grain power allocation. 
\item Finally, we demonstrate the necessity and utility of component-awareness for concurrent kernel executions in ML, a critical paradigm prevalent in ML. 
\item We conclude with recommendations on how component-aware software-hardware co-design can extract energy efficiency from modern GPUs.
\end{enumerate}
\putsec{meth}{Methodology}

\begin{table}[t]
    \centering
    \caption{Evaluated ML GEMMs \& communication collectives (all-gather).}
    \label{tab:gemm_comm_sizes}
    
    \setlength{\tabcolsep}{12pt} 
    \renewcommand{\arraystretch}{1}
    
    \begin{tabular}{lll}
        \toprule
        % GEMM & Source & Comm. & Classification \\
        \multicolumn{1}{c}{GEMM} & \multicolumn{1}{c}{Source} & \multicolumn{1}{c}{Comm.} \\
        \midrule
        (16384,106496,8192) & LLaMA-405B & 4    GB \\
                            &            & 26.5 GB \\
        % \midrule 
        (18432,16384,16384) & LLaMA-405B & 1.5  GB \\
                            &            & 3.5  GB \\
        % \midrule
        (8192,57344,8192)   & LLaMA-70B  & 7    GB \\
        % \midrule
        (8192,8192,10240)   & Synthetic  & 160  MB \\
        \bottomrule
    \end{tabular}
    \vspace{-\baselineskip}
\end{table}

\noindent\textbf{Power Measurement Methodology:} We use FinGraV~\cite{finGraV2025}, a state-of-the-art fine-grain power measurement methodology to obtain the power profiles in this work. FinGraV is cognizant of underlying averaging behavior of power loggers, tackles CPU-GPU time synchronization, execution/power variation and more and consequently leads to high fidelity power/energy measurement~\cite{SCnvSMI24}. FinGraV deploys the dual strategies of executing a given kernel repeatedly in a single run and stitches multiple such runs to realize its power profiles. This helps tackle execution/power variation, noise in power measurements and helps avoid steep errors (as high as 80\%) in power measurements as faced by prior works. Additionally, FinGraV also provides a component-level view of power, that is, a breakdown of the GPU power into MI300X GPU components: XCD, IOD and HBM. Finally, we will focus on relative power data only (not absolute power numbers). As we make evident throughput the paper, using relative power data does not preclude extraction of interesting insights.

\noindent\textbf{Experimental Setup:} We focus on the AMD Instinct MI300X Platform consisting of an 8$\times$ MI300X node with a fully-connected topology. Further, we make all our measurements with default/out-of-box power management infrastructure available on the GPU.
Using this setup we study two primary bottlenecks in ML workloads, namely, general matrix multiplication/GEMM and multi-GPU communication collectives. These two operations together occupy a substantial portion of ML execution time~\cite{totc} and as such analyzing/optimizing for them stands to deliver strong returns. That said, attention computations~\cite{vaswani2023attentionneed} are another important bottleneck for ML and we leave analyzing them for future work. Finally, with regards to multi-GPU ML communication collectives, we focus in this work on widely deployed all-gather collective but other collectives lead to similar profiles~\cite{finGraV2025}. 
While these workloads involve multi-GPU communication, all reported measurements correspond to a single GPU within the node.

\noindent\textbf{ML Operators \& Execution Patterns Under Study:} We list the evaluated GEMMs and all-gather sizes in \tabref{tab:gemm_comm_sizes} along with the source ML models whose execution manifest these GEMM sizes. Further, we study these operators in standalone and concurrent execution scenarios. Both of these execution scenarios manifest in realistic ML inference and training deployments. For standalone, only a single kernel/operator is running on the entirety of the GPU. For concurrent execution, both GEMM and all-gather are executed concurrently sharing available resources. Note that as GPUs get more powerful (higher compute and memory throughput), increasingly, concurrency is often deployed to extract better utilization from the GPU. This is evident as many ML algorithmic choices such as FSDP~\cite{zhao2023fsdp} execute more than a single ML operator on the GPU concurrently. 
\putsec{compow_standalone}{\PNAME Benefits for Standalone Execution}

\begin{figure}[t]
    \centering
    \includegraphics[width=0.8\linewidth]{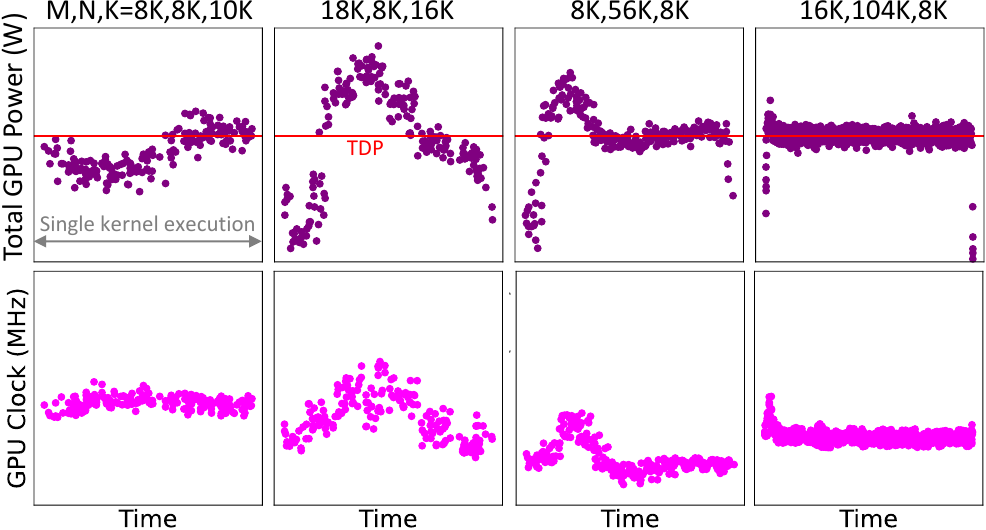}
    \caption{Total power (top) and GPU clock frequency (bottom) for GEMMs.}
    \label{fig:sln_gemm_total_clock}
    \vspace{-\baselineskip}
\end{figure}

In this section, we focus on how component-awareness (\PNAME) can help extract additional energy efficiency during standalone execution of ML operators on GPUs. As discussed in \secref{meth}, we focus on two key ML operators which are critical to performance for both ML training and inference - namely, general matrix multiplication (GEMM) and communication collective (all-gather). Note that, while prior work~\cite{finGraV2025} studied GEMM and all-gather standalone power behavior, we instead focus on characteristics relevant to how component-awareness can optimize power and energy efficiency.

\begin{figure}[t]
    \centering
    \includegraphics[width=0.8\linewidth]{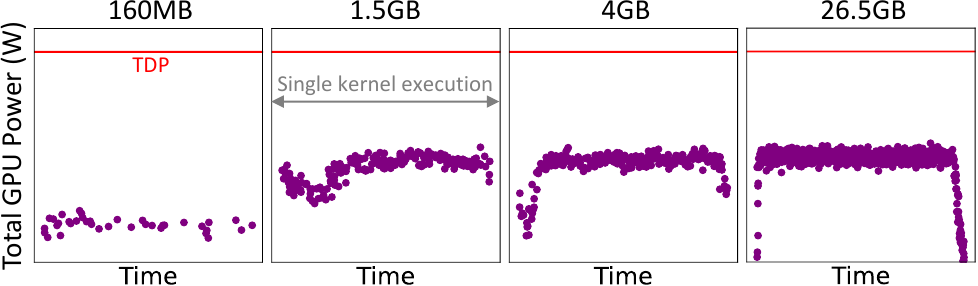}
    \caption{Total power for all-gather.}
    \label{fig:sln_comm_total}
    \vspace{-\baselineskip}
\end{figure}

\putssec{sln_total_gpuclock}{Total Power \& GPU Clock}
We depict the total GPU power and GPU clock for GEMMs under study in \figref{fig:sln_gemm_total_clock}. As depicted, GEMMs can be power constrained on GPUs and further the GPU clock is commensurately clamped to stay within the GPU TDP. In contrast, all-gather total power depicted in \figref{fig:sln_comm_total} shows that communication is not power-constrained. 

\begin{figure}[t]
    \centering
    \includegraphics[width=0.8\linewidth]{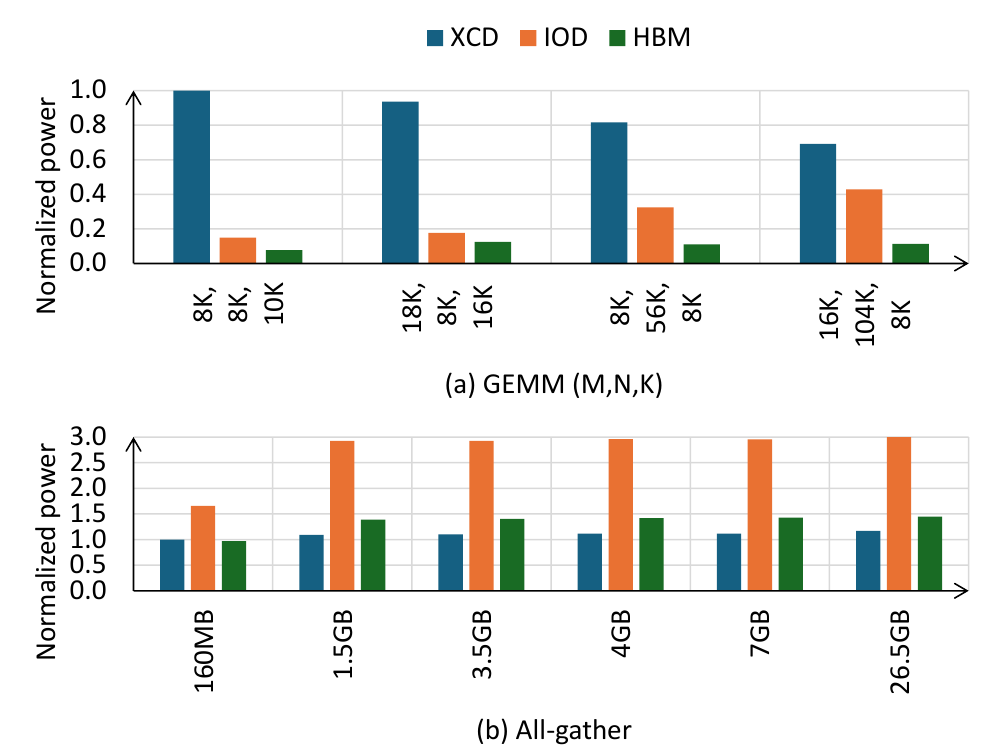}
    \caption{Component-level power breakdown for (a) GEMMs and (b) all-gather. In (a), values are normalized to the XCD power of an (8K × 8K × 10K) GEMM. In (b), values are normalized to the XCD power of a 160M all-gather.}
    \label{fig:sln_comp_jostle}
    \vspace{-\baselineskip}
\end{figure}

\begin{figure}[t]
    \centering
    \includegraphics[width=0.75\linewidth]{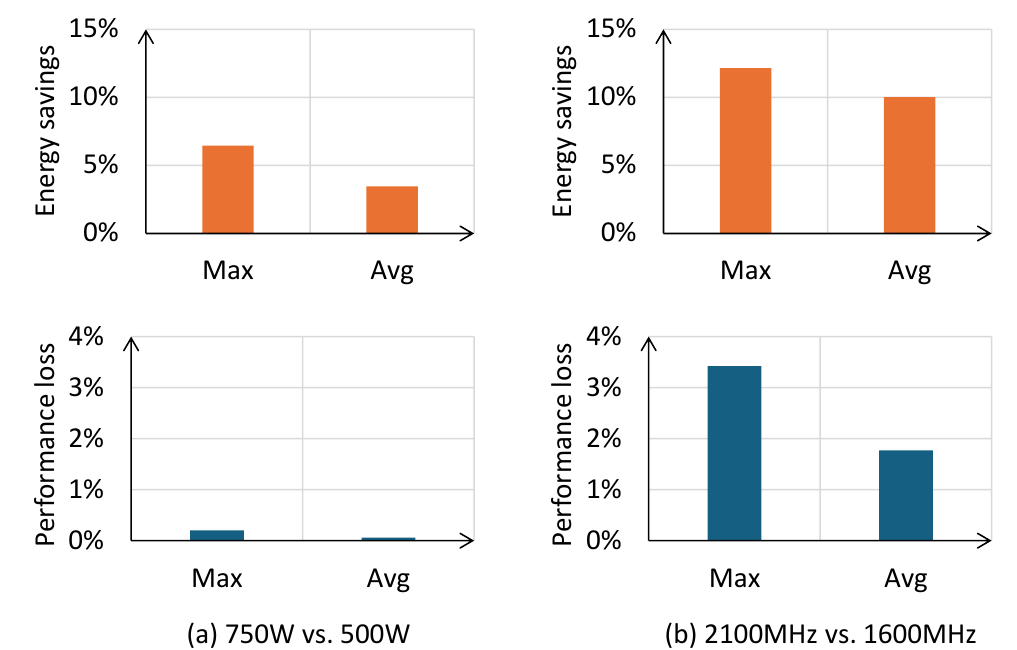}
    \caption{\PNAME optimization for all-gather using general power capping and targeted frequency capping.}
    \label{fig:sln_comm_optimization}
    \vspace{-\baselineskip}
\end{figure}

\begin{figure}[t]
    \centering
    \includegraphics[width=0.8\linewidth]{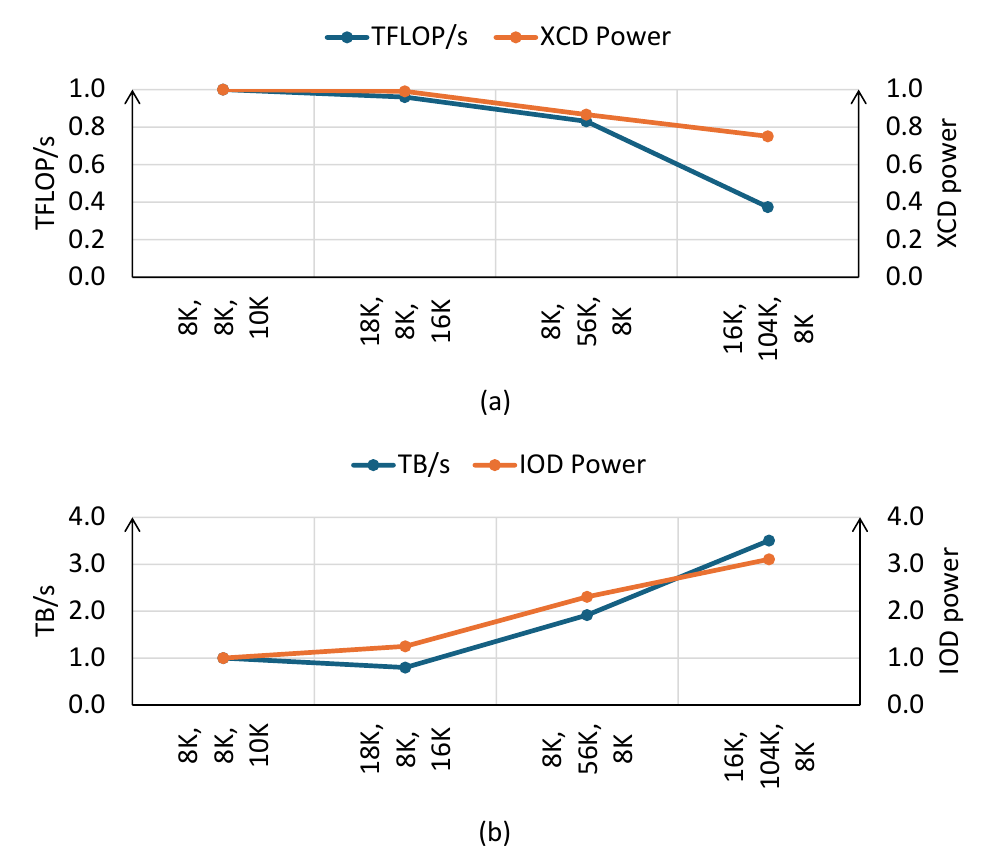}
    \caption{\PNAME optimization potential for GEMM. 
    In (a) values are normalized to the TFLOP/s, XCD power, while in (b) they are normalizes to TB/s, and IOD power of an (8K × 8K × 10K) GEMM.}    
    \label{fig:sln_gemm_optimization}
    \vspace{-\baselineskip}
\end{figure}

\putssec{sln_comp_jostle}{Component-level Power Jostle}
We show component-level breakdown (average power) for GEMMs and all-gather in \figref{fig:sln_comp_jostle}. As depicted, for GEMMs, while XCD is the dominant component (houses matrix cores, L2 cache, etc.), as arithmetic intensity drops (left-to-right), data-movement increases and consequently, IOD component (houses LLC, interface to in-package HBM stacks) starts to increase. Given a fixed power budget for the GPU, this causes power manager to take power from the XCD component, thus leading to a negative correlation between XCD and IOD components. Note that, we use measured arithmetic intensity (measured operations and bytes transferred) which differs from theoretical arithmetic intensity which does not factor underlying GEMM implementations and resultant reuse.

In contrast, for all-gather, as data-movement is dominant, its IOD (and then HBM) component dominates. That is, all-gather comprises of each GPU sending a buffer to its peers and primarily involves loads/stores. Consequently, XCD component is not stressed for all-gather. That said, as all-gather is not power-constrained, component-level power jostling is not present in all-gather and negative correlation between XCD and IOD as in GEMM is not observed. 

\putssec{sln_comm_optimization}{\PNAME Benefits for All-gather}

As we discussed above, different ML operators, pursuant to their algorithmic characteristics, stress different GPU components. As such, a targeted strategy cognizant of this can extract higher energy efficiency without commensurate performance loss. To show this, we evaluate two existing knobs available in modern GPUs to control power allocation, namely, component-agnostic GPU-level power capping and component-aware frequency capping. For the latter, MI300X GPU allows frequency capping of the XCD-level clock (termed GPU clock). 

We tried multiple power capping thresholds gated by the max power consumption we observed across different all-gather kernels and depict the maximum and average energy savings and performance loss in \figref{fig:sln_comm_optimization}. Here we show that with component-agnostic power capping (\figref{fig:sln_comm_optimization}a), an average energy savings of 3.47\% is possible at about 0.06\% performance loss. On the other hand, a more component-aware strategy, which exploits that fact that XCD is not the most stressed component for all-gather, and consequently caps the XCD frequency only, delivers on average 10.13\% energy savings at 1.36\% performance loss (\figref{fig:sln_comm_optimization}b). This demonstrates that component-awareness can lead to higher returns in terms of energy efficiency. We also studied combining frequency capping and power capping and observe that while it can deliver additional energy savings, it also leads to steeper performance loss (up to 7\%). 

\putssec{sln_gemm_potential}{\PNAME Benefit Potential for GEMM}

Unlike all-gather communication collective, which is not power-constrained and lends to coarser-grain component-level power control, GEMMs being power-constrained require a more nuanced approach to extracting higher energy efficiency. Further, commercial GPUs do not expose all the component-level knobs (e.g., frequency capping for every major component) that would allow a more careful study in this regard.  Given these challenges we explore in this section the potential for \PNAME to enable superior power allocation for GEMMs. 

To demonstrate this, we depict in \figref{fig:sln_gemm_optimization}a normalized compute utilization (tera floating point operations per second) and XCD power and in \figref{fig:sln_gemm_optimization}b we depict the normalized LLC bandwidth and IOD power for GEMMs under study. We observe a good (yet loose) correlation between key utilization metrics (compute utilization, bandwidth utilization) and observed power utilization at component-level. This, we believe, opens up the possibility of intelligent phase-level power allocation between components based on tracking of such utilization metrics at fine-granularity. That is, based on phase-level software hints, a component-aware power manager can provision better power jostling between components and more intelligently direct power within a single GPU. 
\putsec{compow_concur}{\PNAME Benefits for Concurrent Execution}

\begin{figure}[t]
    \centering
    \includegraphics[width=0.9\linewidth]{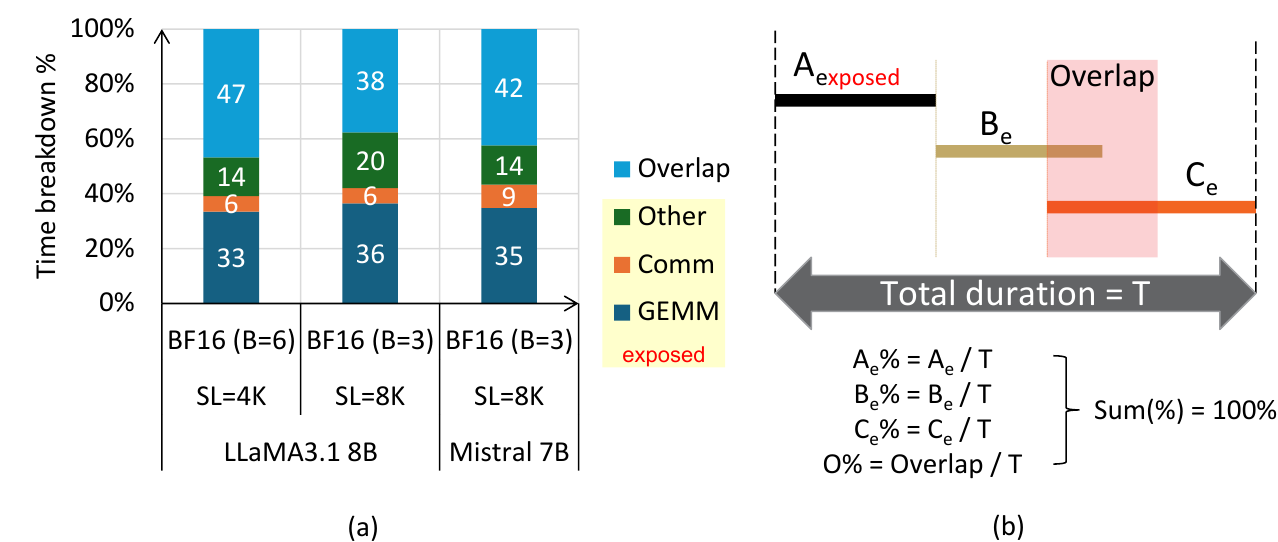}
    \caption{(a) Importance of concurrent execution in realistic training scenarios (datatype=BF16, B/batch-size, SL/sequence-length= 4K or 8K). (b) Exposed and overlapped execution accounting.}
    \label{fig:conc_train_profile}
    \vspace{-\baselineskip}
\end{figure}

\begin{figure}[t]
    \centering
    \includegraphics[width=0.9\linewidth]{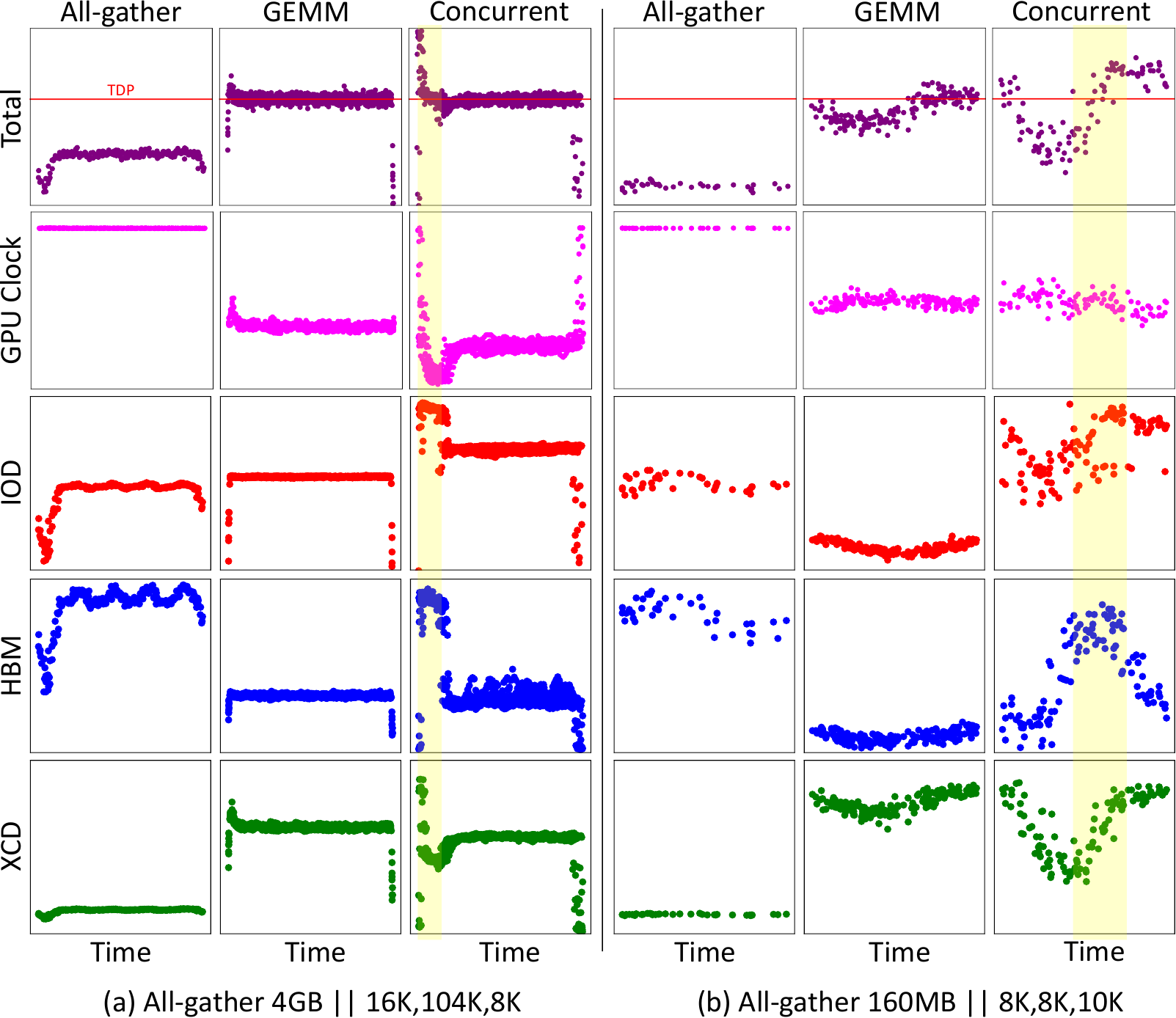}
    \caption{Component-level power jostle for concurrent GEMM/all-gather execution.}
    \label{fig:conc_total_clock_aid_hbm_xcd}
    \vspace{-1.5\baselineskip}
\end{figure}

In this section, we focus on how component-awareness (\PNAME) can extract additional energy efficiency during concurrent execution of multiple ML operators on GPUs. We focus on the most widely deployed concurrent execution scenario in ML: GEMM in concurrence with communication collective (all-gather). 

\putssec{concur_imp}{Concurrency is Prevalent and Critical in ML}

We first depict the importance of concurrent kernel execution in ML in \figref{fig:conc_train_profile}a. We depict here the execution time breakdown for training across different ML models (LLaMA, Mistral) and across different configurations exercised (varying sequence-length and batch-size). We breakdown the execution time into GEMM-only (standalone), communication-only (standalone), other operators (standalone) and finally overlap (concurrent kernels). The methodology we used to calculate overlap execution is depicted in \figref{fig:conc_train_profile}b. We observe here that up to 47\% of overall execution is spent on concurrent kernel execution. That is, more than one kernel is active on the GPU at the same time. Further, most of this overlapped execution time is expended in GEMM/communication concurrency, which is what we focus on in this work. 

\putssec{concur_imp}{Component-level Power Jostle}
We depict total power, IOD/HBM/XCD power in \figref{fig:conc_total_clock_aid_hbm_xcd}, side-by-side for all-gather (standalone), GEMM (standalone), all-gather/GEMM (concurrent) for two scenarios under study. We depict two all-gather/GEMM concurrent execution scenarios for space reasons but the observations we discuss next hold for other concurrent scenarios we studied (listed in \tabref{tab:gemm_comm_sizes}). 

We make several observations here. First total power is stressed more in concurrent execution in comparison to standalone GEMM or all-gather (we highlight portion of execution where two kernels were active in the figure), which is expected. Next, we observe that the IOD component for concurrent execution is higher than either standalone GEMM or all-gather, as data-movement in both GEMM and all-gather, in concurrence, stresses IOD. We observe this behavior for HBM component as well (to a lesser degree). Finally, given a fixed power budget for the GPU, as was observed for standalone GEMM (\ssecref{sln_comp_jostle}), XCD component compensates for higher power draw in IOD. Overall, even for concurrent execution, we observe component-level power jostling within the GPU. 

\putssec{concur_optimization}{\PNAME Benefits for Concurrent Kernel Execution}

\begin{figure}[t]
    \centering
    \includegraphics[width=0.9\linewidth]{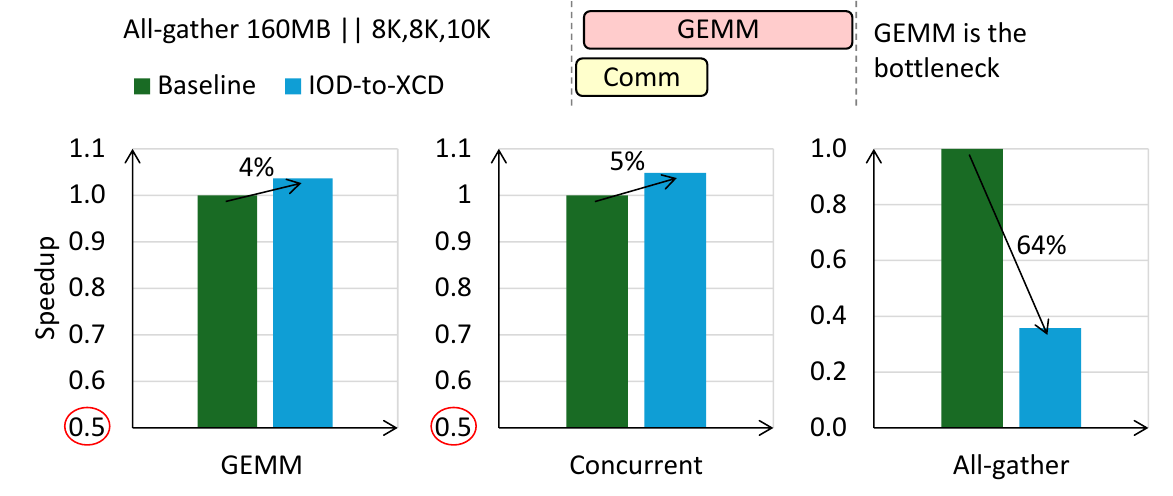}
    \caption{Emulation of IOD-to-XCD power reallocation for concurrent execution.}
    \label{fig:conc_optimization}
    \vspace{-1\baselineskip}
\end{figure}

Concurrent executions, due to their power-constrained nature, face similar challenges to realizing component-level power awareness as GEMMs (\ssecref{sln_gemm_potential}) given the lack of component-level knobs (e.g., frequency capping for every major component) in GPUs. So, to demonstrate the benefits of \PNAME, we emulate different component-level power (re-)allocation by harnessing a feature in MI300X GPU that allows specifying the number of GPU cores to be allocated to each kernel during concurrent execution. 

Specifically, as IOD is stressed during concurrent execution causing XCD power to drop, we aim to allocate more power to XCD and assess the effect on overall performance. To attain this, we emulate additional power allocation from IOD-to-XCD by constraining the GPU CUs allocated to all-gather kernel (which stresses the IOD component the most). Since such power re-allocation will be beneficial where GEMM is the bottleneck, we pick a scenario with GEMM as the bottleneck and depict the resulting performance in \figref{fig:conc_optimization}. We show that such power reallocation leads to a 4\% improvement in GEMM performance and 5\% improvement in concurrent execution by trading off all-gather performance (not on critical path). As such, based on software hints on criticality of an operation during concurrent execution, a component-aware power manager can reallocate power within a single GPU to benefit the operation on the critical path. 
\putsec{codesigh}{Hardware-Software Co-design for \PNAME}

We discuss in this section some possible hardware-software co-design options that can realize \PNAME power management. 

\noindent\textbf{Component-affinity:} In order to aid power manager in making component-aware decisions, software can provide component affinity hints. Using such hints, power manager can deploy targeted power optimizations. An example of this is targeted frequency capping that is useful for communication collectives kernels (\ssecref{sln_comm_optimization}). In some cases, runtime can infer component affinity hints. For example, in case of GEMMs, arithmetic intensity can aid in inferring component affinity; low arithmetic intensity GEMMs can stress data-movement more in comparison to high arithmetic intensity GEMMs (\ssecref{sln_comp_jostle}). Similarly, for ML training, operations are repeated over multiple queries or epochs. In such cases, component affinities for operations can be learned online in some iterations. Similarly, profiling can be employed to learn component affinities offline and provided to the power manager. In some additional cases, such as GEMMs, kernel-level affinity hints may not be coarse (\ssecref{sln_gemm_potential}). In such cases, fine-grain hints demarcating phases in kernel (intra-kernel) can potentially aid in providing component affinities. 

\noindent\textbf{Operation criticality:}  While component-level affinities can aid in targeted power management decisions, concurrent execution can require appropriate combination of such affinities for multiple kernels. In such cases, software hints on operation criticality can aid in prioritizing component affinity for the operation most critical for forward progress (\ssecref{concur_optimization}). In absence of such criticality hints, simple combination (e.g., addition) of component-level affinities can aid the power manager in sloshing power to the appropriate component.  
\putsec{related}{Related Work}

\noindent\textbf{Power Measurement:}
Prior work relies on vendor telemetry interfaces~\cite{amdsmi,nvsmi} and abstractions that standardize power measurement, such as Variorum~\cite{variorum,sandia_standard}, while other approaches use external power meters for higher sampling rates and direct system-level measurement~\cite{pwr_meter}. Additional efforts employ models and simulators for fine-grained estimation~\cite{accelwattch,gpujoule,adhinarayanan_iiswc16,IPDPS14,HPCA15}. In this work, we use FinGraV~\cite{finGraV2025} to obtain high-fidelity, fine-grained power profiles directly from native telemetry.

\noindent\textbf{Standalone Execution:}
Optimizations for standalone execution have evolved from coarse-grained frequency, voltage, and power/frequency capping analysis~\cite{Price_2015,WORKS19} 
to fine-grained, framework-specific tuning~\cite{CCGrid20,ZeusYou23}.
Recent approaches replace reactive heuristics with predictive learning~\cite{DynPowerHPCA2017,WangDRLCAP2024}, while LLM-specific methods exploit distinct iteration-level behaviors~\cite{throttLLem2025}
or data composition~\cite{gregersen_arxiv24}
to optimize power.

\noindent\textbf{Concurrent Execution:}
Concurrent execution improves resource utilization but complicates power management. Early schedulers indirectly optimized energy via QoS and reduced starvation~\cite{prophet_2017,antman_2020}, while later works targeted context-switching overheads~\cite{pipeswitch_2020}. Recent works demonstrate energy savings from spatial partitioning~\cite{zhang2023migperfcomprehensivebenchmarkdeep}, develop models for partition-level power attribution~\cite{migpowermodel2025}, and leverage DMA engines to minimize contention and additionally also deliver improved energy efficiency~\cite{conccl_2025,pati2026}.

\noindent\textbf{Cluster-level Execution:}
Optimization at the cluster-level shifts to holistic scheduling, treating power as a scarce resource for job packing~\cite{gu2023energyefficientgpuclustersscheduling} or incorporating carbon intensity when scheduling~\cite{patterson2021carbon}. To address distributed training inefficiencies, recent works leverage pipeline parallelism bubbles~\cite{envpipe_usenix23,perseusChung24}. For LLMs, specialized strategies emerged to disaggregate prefill-decode phases~\cite{splitwise24}, or safe power oversubscription~\cite{msft_polca}.

\noindent\textbf{Integrating Component‑Level Power Controls:} Component-aware power management in \PNAME unifies these perspectives by providing finer‑grained controls that benefit standalone, concurrent, and cluster-level strategies we discussed above. At the device level, \PNAME enables more efficient use of the local power budget. Under concurrency, \PNAME allows power to be reclaimed or sloshed when concurrent workloads compete for power. At the cluster level, exposing these controls to the scheduler provides new opportunities for power-aware job placement and more efficient global power coordination.
\putsec{conclusion}{Conclusion}
We observe in this work that as modern GPUs are comprised of integrated components, component-awareness when it comes to power management, termed \PNAME, can extract additional energy efficiency from GPUs. We demonstrate the efficacy of \PNAME for different operators that manifest as bottlenecks in ML (GEMM, communication collectives), under various execution scenarios (standalone, concurrent) and show that \PNAME can deliver higher energy efficiency (10\%) and improved performance (5\%). As ML-driven intelligence, fueled by GPUs gets ever more pervasive, intra-GPU power allocation mechanisms like \PNAME stand to play an important role in optimizing GPU-level power which we believe can improve efficacy of cluster-level power allocation/optimizations. 
\vspace{-0.5\baselineskip}
%%%%%%% -- PAPER CONTENT ENDS -- %%%%%%%%
\begin{credits}
\subsubsection{\ackname} The authors thank Nuwan Jayasena and the anonymous EESP reviewers for helping improve the paper. AMD, the AMD Arrow logo, AMD Instinct, AMD Infinity Cache, and combinations thereof are trademarks of Advanced Micro Devices, Inc. Other product names used in this publication are for identification purposes only and may be trademarks of their respective companies.
\end{credits}

\bibliographystyle{splncs04}
\bibliography{ref}

\end{document}